\documentclass[sn-mathphys]{sn-jnl}

\jyear{2021}%
\usepackage{enumerate}
\usepackage{natbib}
\usepackage{mathrsfs}
\usepackage{amsfonts}
\usepackage{subfigure}
\usepackage{amsthm,amsmath,amssymb}
\usepackage{mathrsfs}
\usepackage{multirow}
\usepackage{url}
\usepackage{physics}

\theoremstyle{thmstyleone}%
\theoremstyle{thmstyletwo}%

\theoremstyle{thmstylethree}%

\begin{document}

\title[Modeling dynamic volatility under uncertain environment]{Modeling dynamic volatility under uncertain environment with fuzziness and randomness}


\author[1]{\fnm{Xianfei} \sur{Hui}}

\author[1]{\fnm{Baiqing} \sur{Sun}}

\author[2]{\fnm{Hui} \sur{Jiang}}

\author[1]{\fnm{Yan} \sur{Zhou}}
\equalcont{Corresponding author:Yan Zhou. zyhittxzz@126.com}

\affil[1]{\orgdiv{School of Management}, \orgname{Harbin Institute of Technology}, \orgaddress{\city{Harbin}, \postcode{150001}, \country{China}}}

\affil[2]{\orgdiv{College of Management and Economics}, \orgname{Tianjin University}, \orgaddress{ \city{Tianjin}, \postcode{300072}, \country{China}}}


\abstract{The problem related to predicting dynamic volatility in financial market plays a crucial role in many contexts. We build a new generalized Barndorff-Nielsen and Shephard (BN-S) model suitable for uncertain environment with fuzziness and randomness. This new model considers the delay phenomenon between price fluctuation and volatility changes, solves the problem of the lack of long-range dependence of classic models. Through the experiment of Dow Jones futures price, we find that compared with the classical model, this method effectively combines the uncertain environmental characteristics, which makes the prediction of dynamic volatility has more ideal performance.}




\maketitle

\section{Introduction}
\label{sec:intro}

In recent years, various uncertain political, health and economic events have occurred frequently, which has not only lead to violent fluctuations in the financial market, but also brought huge economic losses and potential investment opportunities. In the quickly changeable market environment, the accurate prediction of asset volatility is a key basic point for asset pricing, economic prediction, risk management and quantitative investment\cite{gradojevic2022unlocking,zhang2020forecasting}. The market, in fact, is a very complex system, which is affected by many internal and external factors and is full of uncertainty\cite{bian2021fuzzy}. The uncertainty of financial market is generally considered to be mainly divided into two parts - randomness and fuzziness \cite{shapiro2009fuzzy,mbairadjim2021fuzzy}. The volatility of financial assets in the market is not directly observable, so it is usually impossible to obtain accurate information on the intensity of the change on asset return in market transactions, which increases the difficulty of volatility estimation \cite{chen2022investor}. Therefore, how to effectively describe the randomness and fuzziness of asset price process in uncertain market environment, build flexible fuzzy stochastic volatility models, and quickly analyze and accurately predict the fluctuation dynamics of asset price are crucial issues that financial technology research should pay attention to.

Considering the fuzziness of asset price volatility, some scholars have applied fuzzy theory to volatility research and achieved fruitful results\cite{pawlowski2021stochastic,nowak2019pricing}. There are two distinct advantages to treating asset prices as fuzzy parameters. One is that the range of price fluctuations could be described more accurately, and the other is that static observational data would be used to describe dynamic price changes\cite{li2018application}. In addition, the use of fuzzy parameters helps to eliminate the errors caused by some unreasonable historical observation data, improve the accuracy of asset return and price fluctuation tracking, and achieve efficiency and operability of the volatility models\cite{de2018pricing}.

In view of the randomness of asset price fluctuation, most studies use stochastic analysis theory to estimate fluctuation \cite{saporito2019calibration,izzeldin2019multivariate}. In the research of stochastic volatility, it is generally assumed that there are jumps (large fluctuations) in asset prices process\cite{yang2019volatility} and the return rate of assets prices obeys a continuous semi-martingale process\cite{yu2021convergence,qin2019nonparametric}. Barndorff Nielsen and Shephard creatively constructed the realized quadratic variation estimator to decompose the continuous fluctuation and jump fluctuation in the realized fluctuation\cite{barndorff2002econometric}. Subsequently, the conditional probability and intensity distribution of jumps occurring in the asset price process constitute an important area of volatility estimation\cite{fallahgoul2020risk,tan2020pricing}.

Barndorff Nielsen and Shephard(BN-S) model, which is a popular stochastic volatility model with L\' evy process, could capture the random behavior of stationary time series and has achieved theoretical success in the research field of nonparametric methods of high-frequency time series \cite{barndorff2001non}. This is one of the most accepted models in finance. Combined with different application scenarios, the previous relevant studies have proposed effective improvements of the basic model, and carried out multi-dimensional research on the stochastic volatility process of asset price, such as jump capture \cite{chen2022probability,gloter2018jump}, fluctuation estimation \cite{james2018stochastic,posedel2021pricing}, pricing \cite{corsaro2019general,chen2021efficient,posedel2021pricing}, and risk management \cite{bakhshmohammadlou2021hedging,baule2021model}.

The data processing classifier based on artificial intelligence approaches provides new technique-based tools for the research of financial volatility prediction in recent years. Among recent studies, the deep learning with unsupervised learning ability shows obvious advantages in modelling and forecasting financial price. Massive high-frequency data retains rich market transaction information to a greater extent, and more accurately records the rapid price change process in the era of big data. The implementation of data-driven techniques promote the efficiency and effectiveness of asset price fluctuation characteristic analysis\cite{li2021role,aras2021improving,samitas2020machine,lyocsa2021stock,gu2020empirical}.

In most of the research on price volatility, stochastic volatility models are constructed by deterministic information, and the fact that there are a lot of fuzzy and uncertain information in the process of price change caused by various emergencies is ignored \cite{chen2022probability,company2019stable}. In the study of dynamic volatility in an uncertain market environment, the classical BN-S models containing a single Ornstein-Uhlenbeck (OU) process \cite{shoshi2021hedging} often fails due to lack of long-range dependence in practical application \cite{awasthi2021first}. There are few studies about volatility estimation in fuzzy random uncertain environment. Although the parameters, such as interest rate, volatility and the price of risk assets in the random volatility model, are usually directly replaced with the corresponding fuzzy numbers in the existing studies, the price process of risk assets is rarely regarded as a fuzzy random process. Compared with replacing the parameters in the formula with fuzzy numbers one by one, it is easier to obtain any level cut set of deterministic parameters in fuzzy random fluctuation by using fuzzy stochastic process \cite{you2021option}.

Considering the fuzziness and randomness in the uncertainty of the financial market, we select the high-frequency Dow Jones Industrial Average (DJIA) futures trading data as research sample with the goal of capturing the deterministic component in random fluctuations, construct a new generalized BN-S model under uncertain environment through the superposition of fuzzy stochastic processes. Three neural network algorithms are used for classification and parameter evaluation. The new model would flexibly describe the actual changes of the dynamic volatility of asset prices without being limited by time, and provide accurate dynamic jump estimation.

The remainder of the paper is organized as follows. A new generalized BN-S model suitable for uncertain environment is introduced in Section 2. Preprocessing and visual analysis of empirical price data of mini-sized DJIA futures are given in Section 3. Section 4 shows the methods and results of parameter estimation. Deep learning algorithms are used to capture the characteristics of price dynamics, predict the deterministic parameter in the fuzzy stochastic volatility model. Finally, Section 5 summarizes the paper.

\section{Methodology}
\label{sec:meth}
\subsection{BN-S Model}
Intra-day-sampled log returns of financial time series have many common stylized features (such as heavy tails, volatility clustering, long-term correlation, etc), which have been successfully captured by the stochastic volatility model with the L\' evy process in the previous literatures. BN-S model is a popular stochastic volatility model with non-Gaussian (OU) process (L\'evy process), which is often used to analyze the random behavior characteristics of asset price process in the research field of non-parametric methods of high-frequency time series. In the BN-S model, asset price dynamics are driven by Brownian motion, and price volatility is described by the L\' evy process. We give a brief review of the original BN-S model.

In a financial market, a risk-free asset with a constant rate of return r and a risk asset (stock) are traded at a fixed horizon time $T$. The BN-S model assumes that the risk asset price process $S=(S_t)_{t \ge 0}$, defined in a filtered probability space $(\Omega, \mathcal{F}, (\mathcal{F}_t)_{0\le t \le T}, \mathbb{P})$, is given by
\begin{equation}
S_t=S_0 \exp(X_t),
\label{(2.1)}
\end{equation}
$\mathbb{P}$ is considered to be log prices, which follow the Black and Scholes type dynamics, with stochastic volatility driven by a non-Gaussian OU process. The log-return $X_t$ is controlled by
\begin{equation}
dX_t=(\mu + \beta \sigma_t^2)dt+\sigma_t d W_t+\rho d Z_{\lambda t},
\label{(2.2)}
\end{equation}
where $\sigma_t$ is the volatility at time $t$. And the parameters $\mu, \beta, \rho$ are constants, $\rho \le 0$. The variance process is given by
\begin{equation}
d\sigma_t^2=-\lambda \sigma_t^2 dt+dZ_{\lambda t}, \quad \sigma_0^2>0,
\label{(2.3)}
\end{equation}
here $\lambda$ is constants, and $\lambda> 0$. For the physical probability measure $\mathbb{P}$, in this model, stochastic process $W= (W_t)$ is a standard Brownian motion. Observe that BN-S model is driven by an incremental L\' evy process of positive mean recovery, and the stochastic process $Z=(Z_{\lambda t})$ is the subordinator (background driving L\' evy process, BDLP), where the processes $W$ and $Z$ are independent, and $(\mathcal{F}_t)$ is a conventional enhancement of the filtering produced by $(W, Z)$.
Solving formula (2.3), $\sigma_t^2$ could be given explicitly by
\begin{equation}
 \sigma_t^2=e^{-\lambda t}\sigma_0^2+\int_{0}^{t} e^{-\lambda (t-s)}\, dZ_{\lambda s},\sigma^2 (0)>0.
\label{(2.4)}
\end{equation}
Obviously, the volatility process $\sigma_t^2$ is constrained by the deterministic function $e^{-\lambda t}\sigma_0^2$, and is strictly positive on $[0, T]$.

Compared with the non-parametric models, the BN-S model is easy to implement mathematically (without assuming that the volatility process obeys a certain distribution, the parameter estimation process is simpler and the estimation efficiency is higher), and could grasp the main characteristics of the realized volatility in the actual market (peak and thick tail, etc.) and the leverage effect of the price itself when describing the asset price change. However, with the continuous research and the development of technique, BN-S model cannot fully adapt to the actual financial market. With the aim of improving the theoretical framework of the model, we consider the price influence conditions in the classical BN-S model, and improve model based on the following two perspectives.

Firstly, improve the description of uncertain market information in original model. As mentioned in the introduction, many literatures have proved that the uncertainty of financial market includes randomness and fuzziness, which have a certain correlation, but can’t replace each other\cite{bian2021fuzzy, anzilli2018pricing, zmevskal2022soft}. It is necessary to consider both randomness and fuzziness so as to comprehensively describe the evolution process of asset price. The classical model only considers the randomness of the price process and ignores the fuzziness. Therefore, in subsection 2.2, we propose a fuzzy random representation of the asset price process, replace the historical price with the corresponding fuzzy number, and treat the asset price process as a fuzzy random process, which are applied to a new generalized BN-S model.

Secondly, solve the problem that BN-S model lacks long-range dependence. Some studies have shown that the superposition of OU processes can effectively improve the long-range dependence of financial random volatility model \cite{barndorff2001superposition}. Considering the asynchronous phenomenon of asset price jump and volatility jump caused by the delay of market information dissemination, some recent literatures constructed the BDLP of BN-S model through two mutually independent L\' evy processes, and achieved the accurate jump capture of crude oil, stock index, option and other financial assets over a long period of time \cite{salmon2021fractional, sengupta2021refinements}. Motived by these facts, in subsection 2.3 we use the superposition of two independent fuzzy L\' evy processes to creatively construct a new generalized BN-S model that could be used in uncertain environment with fuzziness and randomness, and analyze the advantages of the new model.

\subsection{Fuzzy-random Represention of the Assets Pricing Process}
The change of asset price is usually a continuous process, while the empirical price we obtained is discrete and static. The observable price sampling data eliminates some effective information about price changes in the sampling interval, and can't accurately describe the asset price process in the sampling interval. Although the application of high-frequency data solves the above difficulties to a certain extent (frequent sampling reduces the neglected effective information in the sampling interval), but there are still more problems. For example, high-frequency data often carry more invalid information, and serious market microstructure noise makes it difficult to guarantee the unbiasedness and consistency of the realized volatility estimate.

Fuzzy theory could be used to solve the above problem. The discrete asset prices we observed naturally have fuzzy attributes. For example, the price of a stock is \$100 at a certain time. \$100 only represents the static price at one point. In fact, the dynamic price at some point should be described as around \$100. This fact is difficult to describe from the perspective of probability theory. The observed price dynamics could be described as an interval number with membership function from the perspective of fuzzy theory.

We consider implementing two steps of asset price fuzzy representation in the study of dynamic volatility estimation. In the construction of generalized BN-S model, fuzzy random variables could be used to replace the random variables in the original model, and the asset price process would be regarded as a fuzzy random process. At the same time, in the data preprocessing, the fuzzy price could be used to describe the uncertain market information, and the historical price would be replaced by the corresponding fuzzy number.

Now give a brief introduction to the properties and calculation rules of fuzzy numbers, triangular fuzzy numbers, fuzzy random variables in fuzzy theory used in the study of dynamic volatility in fuzzy random uncertain environment.

Suppose that there is a fuzzy set $\tilde{a}$ on domain $S$, where $x$ is any element on all real numbers set $\mathbb{R}$, $\mu_{\tilde{a}}(x) \in [0,1]$ is the membership function of $\tilde{a}$. Membership (confidence) $\alpha$ is usually expressed as $\tilde{a}_\alpha =\{ x:\mu_\alpha \textgreater \alpha \}$. The $\alpha^{-}$ level set of $\tilde{a}$ is composed of all elements whose membership degree to $\tilde{a}$ is not less than $\alpha$. Fuzzy number $\tilde{a}_{\alpha}$ is defined as $\tilde{a}_{\alpha}=\mu_a(x)=[\tilde{a}_{\alpha}^{L},\tilde{a}_{\alpha}^{U}], \alpha \in [0,1]$, where $\tilde{a}_{\alpha}^{L}$ is used to describe the degree of support for $x \in a$, which is the lower bound of membership; and $\tilde{a}_{\alpha}^{U}$ represents the upper bound of membership, describing the degree of opposition to $x \in a$.

In this paper, fuzzy price is used to describe the set of all prices in the sampling interval instead of the real closing price at a certain point. Fuzzy price is the expectation of an interval number. The lower limit of the interval is the lowest price in the sampling period, and the upper limit of the interval is the highest price in the sampling period. Of course, the closing price is also critical. The membership function of fuzzy price interval represents the membership degree of closing price to all price sets.

One of the classical expressions of fuzzy numbers is triangular fuzzy number\cite{al2021ishita}. The membership function of triangular fuzzy number was defined by

$$
\mu_a(x)=\left\{
\begin{array}{lcr}
0, x \leq \tilde{a}_{\alpha}^{l} \\
\dfrac{x-\tilde{a}_{\alpha}^{l}}{\tilde{a}_{\alpha}^{m}-\tilde{a}_{\alpha}^{l}},\tilde{a}_{\alpha}^{l} \leq x \leq \tilde{a}_{\alpha}^{m} \\
\dfrac{\tilde{a}_{\alpha}^{u}-x}{\tilde{a}_{\alpha}^{u}-\tilde{a}_{\alpha}^{m}},\tilde{a}_{\alpha}^{m} \leq x \leq \tilde{a}_{\alpha}^{u}\\
0, \tilde{a}_{\alpha}^{u} \leq x
\end{array}
\right.
0 \leq \tilde{a}_{\alpha}^{l} \leq \tilde{a}_{\alpha}^{m} \leq \tilde{a}_{\alpha}^{u}, 0 \leq \alpha \leq 1.
$$
This function is a continuous convex function, which is composed of linear non decreasing function and linear non increasing function. The real number $\tilde{a}_{\alpha}^{m}$ is called the core value of triangular fuzzy number $\tilde{a}$. The real number $\tilde{a}_{\alpha}^{l}$ and $\tilde{a}_{\alpha}^{u}$ are respectively called the left and right spreads of $\tilde{a}$, represent the lower and upper limits of membership. And their difference shows the fuzzy degree of $\tilde{a}$.

The application of triangular fuzzy numbers is considered suitable for accurately describing the fuzzy properties of asset prices. We use the closing, the highest, and the lowest price as the core value, left and right spreads of the triangular fuzzy price. Compared with the price at sampling time point, which is concerned by previous stochastic volatility models, the triangular fuzzy price not only contains the price information of the sampling time point, but also describes the extent and distribution of dynamic price changes.

In fuzzy stochastic volatility models, fuzzy price calculations are inevitable. We list some basic arithmetic rules for triangular fuzzy numbers.

If $\tilde{a}_{\alpha}^{l}=\tilde{a}_{\alpha}^{m}=\tilde{a}_{\alpha}^{u}$, the fuzzy set $\tilde{a}$ degenerates into real number $\tilde{a}_{\alpha}^{m}$. In other words, the fuzzy price is equal to the historical closing price if the asset price did not fluctuate during the sampled period. The $\alpha^{-}$ level set of $\tilde{a}$ is $\tilde{a}_{\alpha}=[\tilde{a}_{\alpha}^{m}-(1-\alpha)\tilde{a}_{\alpha}^{l},\tilde{a}_{\alpha}^{m}+(1-\alpha)\tilde{a}_{\alpha}^{u}]$, where $\tilde{a}_{\alpha}^{L}=\tilde{a}_{\alpha}^{m}-(1-\alpha)\tilde{a}_{\alpha}^{l}$ and $\tilde{a}_{\alpha}^{U}=\tilde{a}_{\alpha}^{m}+(1-\alpha)\tilde{a}_{\alpha}^{u}$. If $\tilde{a}_{\alpha}^{L}$ and $\tilde{a}_{\alpha}^{U}$ are integrable, the expectation of $\tilde{a}$ is given by $E(\tilde{a}_{\alpha})=\frac{(1-\eta)\tilde{a}_{\alpha}^{l}+\tilde{a}_{\alpha}^{m}+\eta \tilde{a}_{\alpha}^{u}}{2}, \eta \in [0,1]$, where the value of $\eta$ is related to the attention of fuzzy boundary. In the calculation of fuzzy price expectations, the value of $\eta$ is influenced by the risk attitude held by market participants. When investors are risk averse or pessimistic about holdings of the underlying asset, they pay more attention to the bottom of the price change. In this case, the lower limit of fuzzy price is the most important. The value of $\eta$ ranges from $0$ to $0.5$ and tends to $0$ with increasing pessimism or risk aversion. On the contrary, the upper limit of fuzzy price is more critical when decision makers are optimistic about the underlying assets or prefer risk. In this case, the value of $\eta$ is between $0.5$ and $1$, and tends to $1$ with the increase of optimism or risk preference. When the decision-maker takes a neutral attitude towards the risk of the underlying asset, $\eta=0.5$. In addition to the investor risk attitudes we discussed, there are other factors that affect the value of $\eta$, such as market trend, investment target, etc. In a rising bull market, the upper bound of the fuzzy price is more important than the lower bound, and the value of $\eta$ is generally between $0.5$ and $1$. In contrast, the value of $\eta$ in a bear market is usually less than $0.5$. Generally, the lower bound of put option price volatility is more concerned than the upper bound, so the value of $\eta$ tends to $0$ as the drop increases. While the value of $\eta$ describing the price dynamics of call options is always bigger than that of put options, and is generally between $0.5$ and $1$.

Note that the result of multiplication and division between real number $\gamma \not= 0$ and triangular fuzzy number is still triangular fuzzy number, and the sum and difference of two triangular fuzzy numbers $\tilde{a}$ and $\tilde{b}$ are also triangular fuzzy numbers. For any membership $\alpha \in [0,1]$,
$$
\begin{array}{lcr}
\gamma \tilde{a}_{\alpha}=[\gamma \tilde{a}_{\alpha}^{l}, \gamma \tilde{a}_{\alpha}^{m}, \gamma \tilde{a}_{\alpha}^{u}], \\
\frac{\gamma}{\tilde{a}_{\alpha}}=[\frac{\gamma}{\tilde{a}_{\alpha}^{l}},\frac{\gamma}{\tilde{a}_{\alpha}^{m}},\frac{\gamma}{\tilde{a}_{\alpha}^{u}}], \tilde{a}_{\alpha} \not= 0, \\
(\tilde{a}+\tilde{b})_{\alpha}=[\tilde{a}_{\alpha}^{l}+\tilde{b}_{\alpha}^{l},\tilde{a}_{\alpha}^{m}+\tilde{b}_{\alpha}^{m}, \tilde{a}_{\alpha}^{u}+\tilde{b}_{\alpha}^{u}], \\
(\tilde{a}-\tilde{b})_{\alpha}=[\tilde{a}_{\alpha}^{l}-\tilde{b}_{\alpha}^{l},\tilde{a}_{\alpha}^{m}-\tilde{b}_{\alpha}^{m}, \tilde{a}_{\alpha}^{u}-\tilde{b}_{\alpha}^{u}].
\end{array}
$$

Under uncertain environment with fuzziness and randomness, the stochastic process in BN-S model could be replaced by fuzzy random process. A stochastic process is a set of random variables that depend on time. Correspondingly, fuzzy random process would be considered as a set of time-dependent fuzzy random variables.

Random variables with fuzzy forms are usually called fuzzy random variables, which have both randomness and fuzziness. Let $(\Omega, \mathcal{F}, \mathcal{P})$ be a probability space with filtration satisfying standard assumptions, $\mathcal{B}$ is the Borel $\sigma^{-}$ field of all real numbers set $\mathbb{R}$, for every $\alpha \in [0,1]$ satisfying $\{(w,x) \in \Omega \times \mathbb{R} \mid \tilde{X}(w)(x) \geq \alpha\} \in \mathcal{F} \times \mathcal{B}$,the fuzzy-number-valued map $\tilde{X}:\Omega \longrightarrow \mathscr{F}$ is called a fuzzy random variable\cite{kwakernaak1978fuzzy}. The $\alpha^{-}$ cut set of fuzzy random variable $\tilde{X}$ is a closed interval of the form $\tilde{X}=[\tilde{X}_{\alpha}^{L},\tilde{X}_{\alpha}^{U}]$, where $\tilde{X}_{\alpha}^{L}$ and $\tilde{X}_{\alpha}^{U}$ are general random variables. If $\tilde{X}_{\alpha}^{L}$ and $\tilde{X}_{\alpha}^{U}$ are integrable, $\forall \alpha \in [0,1]$, the expectation $E(\tilde{X})$ of $\tilde{X}$ could be written out as
$$
E(\tilde{X})(x)= \mathop{sup}\limits_{\alpha \in [0,1]} min\{\alpha, l_{E{X}_{\alpha}}{x}\}, x\in \mathbb{R},
$$
$$
E(\tilde{X})_{\alpha}=[\int_{\Omega}^{}\tilde{X}_{\alpha}^{L} dP, \int_{\Omega}^{}\tilde{X}_{\alpha}^{U} dP]=[E(\tilde{X}_{\alpha}^{L}),E(\tilde{X}_{\alpha}^{U})], \alpha \in[0,1],
$$
where $l_{A}(x)$ is indicator function (if $x \in A, l_{A}(x)=1$, otherwise $l_{A}(x)=0$). The variance of $\tilde{X}$ is defined as
$$
D\tilde{X}=Cov(\tilde{X},\tilde{X})=\frac{1}{2}\mathop{\int}\limits_{0}^{1}(D(\tilde{X}_{\alpha}^{L})+D(\tilde{X}_{\alpha}^{U}))d\alpha, \alpha \in [0,1].
$$
For two fuzzy random variables $\tilde{X}$ and $\tilde{Y}$, the variance of their sum is
$$
D(\tilde{X}+\tilde{Y})=D\tilde{X}+D\tilde{Y}+2Cov(\tilde{X},\tilde{Y}),
$$
and their covariance is
$$
\begin{small}
Cov(\tilde{X},\tilde{Y})=\frac{1}{2}[E(\tilde{X},\tilde{Y})-E\tilde{X}-E\tilde{Y}]
=\frac{1}{2}\mathop{\int}\limits_{0}^{1}(Cov(\tilde{X}_{\alpha}^{L},\tilde{Y}_{\alpha}^{L})+Cov(\tilde{X}_{\alpha}^{U},\tilde{Y}_{\alpha}^{U}))d\alpha, \alpha \in [0,1].
\end{small}
$$
Their correlation coefficient is
$$
\rho(\tilde{X},\tilde{Y})=\frac{Cov(\tilde{X},\tilde{Y})}{\sqrt{D\tilde{X},D\tilde{Y}}},
$$
with both $D\tilde{X}$ and $D\tilde{Y}$ are not $0$.

Applying the above fuzzy theory to the stochastic volatility model of risk asset price process, a fuzzy BN-S model was got. In this model, fuzzy stochastic price $\tilde{S}_t$ ,fuzzy log-return $\tilde{X}_t$ and fuzzy variance $\tilde{\sigma}_t^{2}$ of the stock or commodity are given by
\begin{equation}
\tilde{S}_t=\tilde{S}_0 exp(\tilde{X}_t),
\label{(2.5)}
\end{equation}
\begin{equation}
d\tilde{X}_t=(\mu+\beta \tilde{\sigma}_t^2)dt+\tilde{\sigma}_t d W_t+\rho d \tilde{Z}_{\lambda t},
\label{(2.6)}
\end{equation}
\begin{equation}
 d\tilde{\sigma}_t^2= -\lambda \tilde{\sigma}_t^2 dt+ d\tilde{Z}_{\lambda t}, \quad \tilde{\sigma}_0^2 >0,
\label{(2.7)}
\end{equation}
where the parameters $\mu, \beta, \rho, \lambda$ are constants, and $\rho \leq 0, \lambda >0$. The process $W_t$ is a standard Brownian motion and $\tilde{Z}_{\lambda t}$ is a subordinator(background driving fuzzy L\' evy process, BDFLP). And $\tilde{\sigma}^{2}=(\tilde{\sigma}_t^{2})$ is strictly positive.

Compared with the original model, fuzzy BN-S model has more advantages. Because the fuzzy random process carries more effective market information, the fuzzy model could describe the dynamic changes of price time series more accurately. In application, fuzzy price is applicable to all decision makers with different risk attitudes in the market. The change of risk attitude coefficient $\eta$ makes the model more flexible. The results of volatility estimation using fuzzy BN-S model have fuzzy attributes. Calculating the expectation value of fuzzy interval number is helpful to improve the accuracy of prediction results.

\subsection{A Generalized BN-S Model}
In this subsection, a new generalized BN-S model suitable for uncertain environment is introduced. Based on the fuzzy BN-S model, given by \eqref{(2.5)}\eqref{(2.6)}\eqref{(2.7)} shown in the previous subsection, the improved new model uses the superposition of two independent fuzzy random processes to fit the dynamic fluctuations of price and volatility.

Refer to a refined BN-S model \cite{shoshi2021hedging}, assuming $\tilde{Z}_{\lambda t}$ and $\tilde{Z}_{\lambda t}^{*}$ to be two independent fuzzy L\' evy subordinators with same (finite) variance. Like the fuzzy random process $\tilde{Z}_{\lambda t}$ independent of $W$ in \eqref{(2.6)}, the new fuzzy L\' evy subordination $\overline{Z}_{\lambda t}$ is defined as
\begin{equation}
 d\overline{Z}_{\lambda t}= \rho' d \tilde{Z}_{\lambda t}+\sqrt{1-\rho'^2} d \tilde{Z}_{\lambda t}^{*}, \quad 0\le \rho' \le 1,
\label{(2.8)}
\end{equation}
where fuzzy random process $\tilde{Z}$ (including $\tilde{Z}_{\lambda t}$ and $\tilde{Z}_{\lambda t}^{*}$) and $\overline{Z}=(\overline{Z}_{\lambda t})$ are positively correlated L\' evy subordinators.

Refer to dynamics of fuzzy stochastic price process of risky assets $\tilde{S}_{t}$ given by \eqref{(2.5)} and \eqref{(2.6)}, and the new fuzzy L\' evy subordination $\overline{Z}_{\lambda t}$ given by \eqref{(2.8)}, fuzzy variance process $\tilde{\sigma}_{t}^2$ in \eqref{(2.7)} could be rewritten as
\begin{equation}
d\tilde{\sigma}_t^2= -\lambda \tilde{\sigma}_t^2 dt+ d\overline{Z}_{\lambda t}, \quad \tilde{\sigma}_0^2 >0,
\label{(2.9)}
\end{equation}
where $E(\tilde{\sigma}_{0}) \not= 0$. It is also a subordinate independent of $W$, and it is similar to $\tilde{Z}$ in \eqref{(2.3)} and $\overline{Z}$ in \eqref{(2.7)}. $Z, \tilde{Z}$ and $\overline{Z}$ are positively correlated L\' evy subordinators.

Given the constant influence of various information in the market, the price and volatility of risky assets may be affected to some extent and result in a series of changes. But these changes often do not happen at the same time. Because of the lag of market response, the jump of volatility and stock price are not always synchronized. As proposed in\cite{sengupta2021refinements}, considering \eqref{(2.8)} on the basis of \eqref{(2.6)}, a new fuzzy log return $\tilde{X}_{t}$ with a convex combination of two independent subordinates $\tilde{Z}_{\lambda t}$ and $\tilde{Z}_{\lambda t}^{(b)}$ will be given by
\begin{equation}
 d\tilde{X}_t= (\mu +\beta \tilde{\sigma}_t^2)dt +\tilde{\sigma}_t d W_t +\rho ((1-\theta)d\tilde{Z}_{\lambda t}+\theta d \tilde{Z}_{\lambda t}^{(b)}),
\label{(2.10)}
\end{equation}
where $\tilde{Z}_{\lambda t}$ and $\tilde{Z}_{\lambda t}^{(b)}$ are fuzzy random processes related to time parameter t and scale parameter $\lambda$. Jump measure $\tilde{Z}_{\lambda t}^{(b)}$ has greater L\' evy intensity than jump measure $\tilde{Z}_{\lambda t}$. $\theta$ is a deterministic parameter,and $\theta \in [0,1]$. Specifically, for $\theta =1$, the original fuzzy stochastic process $\tilde{Z}_{\lambda t}$ will be replaced by $\tilde{Z}_{\lambda t}^{(b)}$. More intensive fuzzy stochastic process $\tilde{Z}_{\lambda t}^{(b)}$ means the big jumps. If there is no big jump, the value of $\theta$ is $0$. $\theta$ would be regarded as the deterministic component for completely stochastic dynamic price processes. And the value of $\theta$ in different periods could be estimated from empirical data. $(1-\theta)d\tilde{Z}_{\lambda t}+\theta d\tilde{Z}_{\lambda t}^{(b)}$ gives a flexible jump measure and accurately describes the dynamic characteristics of asset price fluctuations. In this case \eqref{(2.9)} will be given by
\begin{equation}
d\tilde{\sigma}_t^2= -\lambda \tilde{\sigma}_t^2 dt+ (1-\theta')d\tilde{Z}_{\lambda t}+\theta^{'} d \tilde{Z}_{\lambda t}^{(b)},\tilde{\sigma}_0^2 >0,
\label{(2.11)}
\end{equation}
where, as $\theta$ in \eqref{(2.10)}, $\theta^{'} \in [0,1]$ is also deterministic. According to the operation rules of fuzzy random variables,  $(1-\theta)d\tilde{Z}_{\lambda t}+\theta d\tilde{Z}_{\lambda t}^{(b)}$ is also a fuzzy L\' evy subordination that is positively correlated with both $\tilde{Z}_{\lambda t}$ and $\tilde{Z}_{\lambda t}^{(b)}$. In this paper, we assume $\theta=\theta^{'}$ for simplicity.

The expressions, given by \eqref{(2.5)}, \eqref{(2.10)} and \eqref{(2.11)},  of conditional distribution characteristic function of logarithmic price returns process in a new generalized BN-S model is derived. The parameter $\theta$ in \eqref{(2.10)} and \eqref{(2.11)}, a deterministic component in a completely random price process, could be captured and predicted from historical data through machine learning and deep learning.

There is evidence that the dynamic process given by the new model contains long-term dependence. Suppose jump measures $J_Z, \tilde{J}_Z$ and $\tilde{J}_Z^{(b)}$ are related to subordinators $Z_{\lambda t}$, $\tilde{Z}_{\lambda t}$ and $\tilde{Z}_{\lambda t}^{(b)}$ respectively, and $J(s)=\int_0^s \int_{\mathbb{R}+}J_Z(\lambda d\tau ,dy)$, $\tilde{J}(s)=\int_0^s \int_{\mathbb{R}+}\tilde{J}_Z(\lambda d\tau ,dy)$, $\tilde{J}^{(b)}(s)=\int_0^s \int_{\mathbb{R}+}\tilde{J}_Z^{(b)}(\lambda d\tau ,dy)$; then for the log-return of BN-S model, fuzzy BN-S model, and generalized BN-S model mentioned above,
\begin{equation}
Corr(X_t,X_s)= \frac{\int_0^s \sigma_\tau^2d\tau +\rho^2J(s)}{\sqrt{(\int_0^t \sigma_\tau^2 d \tau + t \rho^2 \lambda Var(Z_{\lambda t}))(\int_0^s \sigma_\tau^2 d \tau + s \rho^2 \lambda Var(Z_{\lambda t}))}}, \quad t>s,
\label{(2.12)}
\end{equation}
\begin{equation}
Corr(\tilde{X}_t,\tilde{X}_s)= \frac{\int_0^s \tilde{\sigma}_\tau^2d\tau +\rho^2\tilde{J}(s)}{\sqrt{(\int_0^t \tilde{\sigma}_\tau^2 d \tau + t \rho^2 \lambda Var(\tilde{Z}_{\lambda t}))(\int_0^s \tilde{\sigma}_\tau^2 d \tau + s \rho^2 \lambda Var(\tilde{Z}_{\lambda t}))}}, \quad t>s,
\label{(2.13)}
\end{equation}
\begin{tiny}
\begin{equation}
Corr(\tilde{X}_t,\tilde{X}_s)= \frac{\int_0^s \tilde{\sigma}_\tau^2d\tau +\rho^2(1-\theta)^2\tilde{J}(s)+\rho^2 \theta^2\tilde{J}^{(b)}(s)}{\sqrt{(\int_0^t \tilde{\sigma}_\tau^2 d \tau +v \rho^2 \lambda ((1-\theta)^2 Var(\tilde{Z}_{\lambda t})+\theta^2 Var(\tilde{Z}_{\lambda t}^{(b)})))(\int_0^s \tilde{\sigma}_\tau^2 d \tau +v \rho^2 \lambda ((1-\theta)^2 Var(\tilde{Z}_{\lambda t})+\theta^2 Var(\tilde{Z}_{\lambda t}^{(b)})}}, \\
\quad t>s.
\label{(2.14)}
\end{equation}
\end{tiny}
The instantaneous variance of log-return in \eqref{(2.12)} and \eqref{(2.13)} are $(\sigma_t^2 + \rho^2 \lambda Var(Z_{1}))dt$ and $(\tilde{\sigma}_t^2 + \rho^2 \lambda Var(\tilde{Z}_{1}^{(b)}))dt$ respectively. $Corr(X_t, X_s)$ will quickly become smaller as t increases for a fixed s. This decrease means that BN-S model and fuzzy BN-S model are affected by time t, which may lead to inaccurate result, that is, the models will fail seriously over a longer time. Note that in \eqref{(2.14)},the variance of log-returns $X_t$ is $\int_{0}^{v}\tilde{\sigma}_\tau^2 d \tau +v\rho^2\lambda((1-\theta)^2Var(\tilde{Z}_{\lambda t})+\theta^2Var(\tilde{Z}_{\lambda t}^{(b)}))$, and the variance of log-returns $X_s$ is given by $\int_{0}^{v}\tilde{\sigma}_\tau^2 d \tau +v\rho^2\lambda((1-\theta)^2Var(\tilde{Z}_{\lambda t})+\theta^2Var(\tilde{Z}_{\lambda t}^{(b)}))$. Unlike the results of the previous two models, the parameter $\theta$ is a dynamic value. For a fixed s in this model, the value of $\theta$ varies, $t$ always has an upper limit, and the model will never fail because $Corr(X_t,X_s)_3$ becomes ``too small". This result proves that the new generalized BN-S model with fuzzy random processes incorporates long-range dependence. Our generalized BN-S model is believed to provide dynamic properties with distinct advantages when applied for modeling dynamic volatility under uncertain environment with fuzziness and randomness.

In conclusion, the new model has three advantages. Under uncertain environment, the fuzzy random variable in the new model is used to describe the fuzziness and randomness in the uncertainty of the financial market at the same time, to improve the utilization of information and the accuracy of results. In view of the jumps from price and volatility are not synchronized, a flexible and effective solution is proposed through less parameter changes. It has the advantage of long-range dependence, could be competent for volatility estimation in long time range, and expands the application range of BN-S volatility models. In the following, the new generalized BN-S model is tried to be used to analyze and predict the dynamic fluctuation of DJIA futures price in a fuzzy and random environment.

\section{Data}
\label{sec:data}
\subsection{Sample Selection} Considering the industry representativeness of DJIA futures listed in the Chicago Board of Trade (CBOT), this paper selects the high-frequency trading data of 1-year mini-sized DJIA futures (YM.CBT) as a sample to carry out dynamic volatility analysis. Its related stock index is DJIA, which is the most widely quoted stock index. The Futures contract months are March, June, September and December, which are listed and traded at all time of the year. Chicago Mercantile Exchange Globex electronic trading is from Sunday to Friday, from 6:00 p.m. to 5:00 p.m. of the next day (US Eastern time), and trading is suspended for 15 minutes from 4:15-4:30 p.m. (US Eastern time). We collected 66403 historical transaction data every 5 minutes in one year from two online data sources (Wind \footnote{\url{https://www.wind.com}} and Yahoo Finance\footnote{\url{https://www.finance.yahoo.com}}). The time range of sample tracking is from 6:00 pm on September 30, 2020 to 5:00 pm on October 1, 2021. According to the sampling frequency of 5 minutes, 273 price data could be observed every day.

\subsection{Data Preprocessing}

With the purpose of estimating the market volatility under uncertain environment with fuzziness and randomness, we implement the generalized BN-S model proposed in Section 2, and preprocess the historical data. Referring to the closing price at the sampling time and the historical price change range in the sampling interval, the triangular fuzzy price expectation, formed by the lowest price, closing price and highest price, would be used as the fuzzy price of the sampling interval.

Now we give the descriptive statistics of the sample of high-frequency mini-sized DJIA futures price after data preprocessing. Price is the key variable concerned in the study of dynamic fluctuation rule. But given the change of the closing price of mini-sized DJIA futures under the 5-minute sampling frequency is very small compared with the closing price itself. Direct observation of price can't fully meet the accuracy requirements of high-frequency fluctuation research, and the small change of price may be ignored. With the view to pay full attention to the change dynamics of price, we consider the percentage change of price as one of the key variables. Table 1 provides descriptive statistics of price, fuzzy price and percentage change of them. When calculating the expectation of fuzzy price interval, $\eta$ represents different investment preferences. The expected value of price fuzzy set and the corresponding fuzzy price change percentage under different values of $\eta$ is showed in the table. Mean, median, minimum, maximum, skewness and kurtosis of eight variables about the mini-sized DJIA futures are reported.

The median of historical price change percentage is 0, indicating that the number of positive and negative fluctuations is basically balanced. In different market situations and investor preferences, the average number of fuzzy prices is distributed between 32521 and 32531.36, which shows that the fuzzy price of assets in the fuzzy random uncertain environment contains more market information than the historical price. For $\eta=0$, the value of fuzzy price is the smallest and the range of negative fluctuation is the largest, while the value of fuzzy price is the largest and the range of positive fluctuation is the largest for $\eta=1$. The skewness of observed price change percentage is greater than 0, and the kurtosis is far greater than 0, which indicate that there is a thick tail phenomenon on the right side and the characteristics of peak of the observation sample. This is consistent with the previous research conclusion of the peak thick tail in the return ratio sequence. The above data features are applicable to the generalized BN-S model with jump process mentioned in Section 2, L\' evy process in the model could describe the dynamic changes of price time series.

\begin{table}[htb]
\center
\vspace{-1.5em}
\caption{Descriptive statistics\label{tab:tabone}}
\tiny
\renewcommand{\arraystretch}{1.5}
\setlength\tabcolsep{1mm}{
\begin{tabular}{|c|c|ccc|c|ccc|}
\hline
\multirow{2}{*}{} & \multirow{2}{*}{\begin{tabular}[c]{@{}c@{}}Close \\ Price\end{tabular}} & \multicolumn{3}{c|}{Fuzzy Price}                                        & \multirow{2}{*}{\renewcommand{\arraystretch}{1} \begin{tabular}[c]{@{}c@{}}Close Price \\ Change \\ Percentage  \end{tabular}} & \multicolumn{3}{c|}{\begin{tabular}[c]{@{}c@{}}Fuzzy Price \\ Change Percentage\end{tabular}} \\ \cline{3-5} \cline{7-9}
                  &                                                                         & \multicolumn{1}{c|}{$\eta=0$}   & \multicolumn{1}{c|}{$\eta=0.5$}  & $\eta=1$    &                                                                                              & \multicolumn{1}{c|}{$\eta=0$}            & \multicolumn{1}{c|}{$\eta=0.5$}       & $\eta=1$            \\ \hline
Mean              & 32526.27                                                                & \multicolumn{1}{c|}{32521}   & \multicolumn{1}{c|}{32526.18} & 32531.36 & 0.00034                                                                                      & \multicolumn{1}{c|}{0.000335}         & \multicolumn{1}{c|}{0.00033}       & 0.000332         \\ \hline
Median            & 33492                                                                   & \multicolumn{1}{c|}{33483.5} & \multicolumn{1}{c|}{33491}    & 33499.5  & 0                                                                                            & \multicolumn{1}{c|}{0.00289118}       & \multicolumn{1}{c|}{0.000721}      & -0.00150515      \\ \hline
Minimum           & 26017                                                                   & \multicolumn{1}{c|}{25985}   & \multicolumn{1}{c|}{26006}    & 26027    & -1.213879                                                                                    & \multicolumn{1}{c|}{-1.2819595}       & \multicolumn{1}{c|}{-1.25}         & -1.22621413      \\ \hline
Maximum           & 35539                                                                   & \multicolumn{1}{c|}{35531}   & \multicolumn{1}{c|}{35535.25} & 35541    & 5.31098617                                                                                   & \multicolumn{1}{c|}{4.99035893}       & \multicolumn{1}{c|}{5.15}          & 5.314522117      \\ \hline
Skewness          & -0.64                                                                   & \multicolumn{1}{c|}{-0.64}   & \multicolumn{1}{c|}{-0.64}    & -0.64    & 12.2762931                                                                                   & \multicolumn{1}{c|}{14.1191821}       & \multicolumn{1}{c|}{18.12}         & 19.7289639       \\ \hline
Kurtosis          & -0.74                                                                   & \multicolumn{1}{c|}{-0.73}   & \multicolumn{1}{c|}{-0.74}    & -0.74    & 1083.42272                                                                                   & \multicolumn{1}{c|}{1391.939}         & \multicolumn{1}{c|}{1794.62}       & 1906.426504      \\ \hline
\end{tabular}
}
\end{table}

\subsection{Analysis}
In this subsection, we use visual analysis to understand the basic rule of mini-sized DJIA futures price time series fluctuation. We show the trend and distribution of fuzzy price, yield and realized volatility over time for $\eta=0.5$.

\begin{figure}[htp]
\begin{center}
\includegraphics[width=3in]{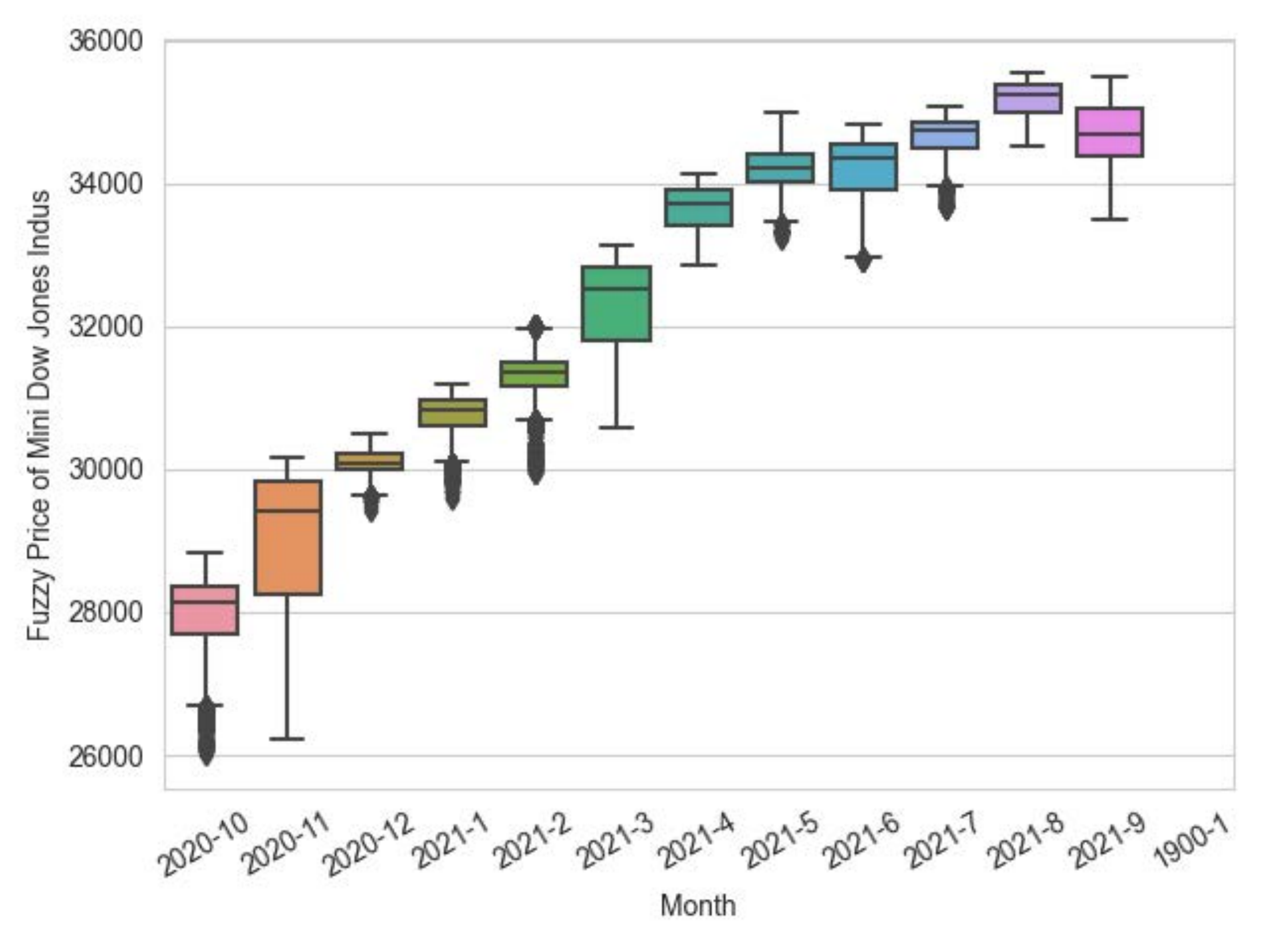}
\end{center}
\caption{Boxplot for fuzzy price. \label{fig:first}}
\end{figure}

\begin{figure}[htp]
\begin{center}
\includegraphics[width=3in]{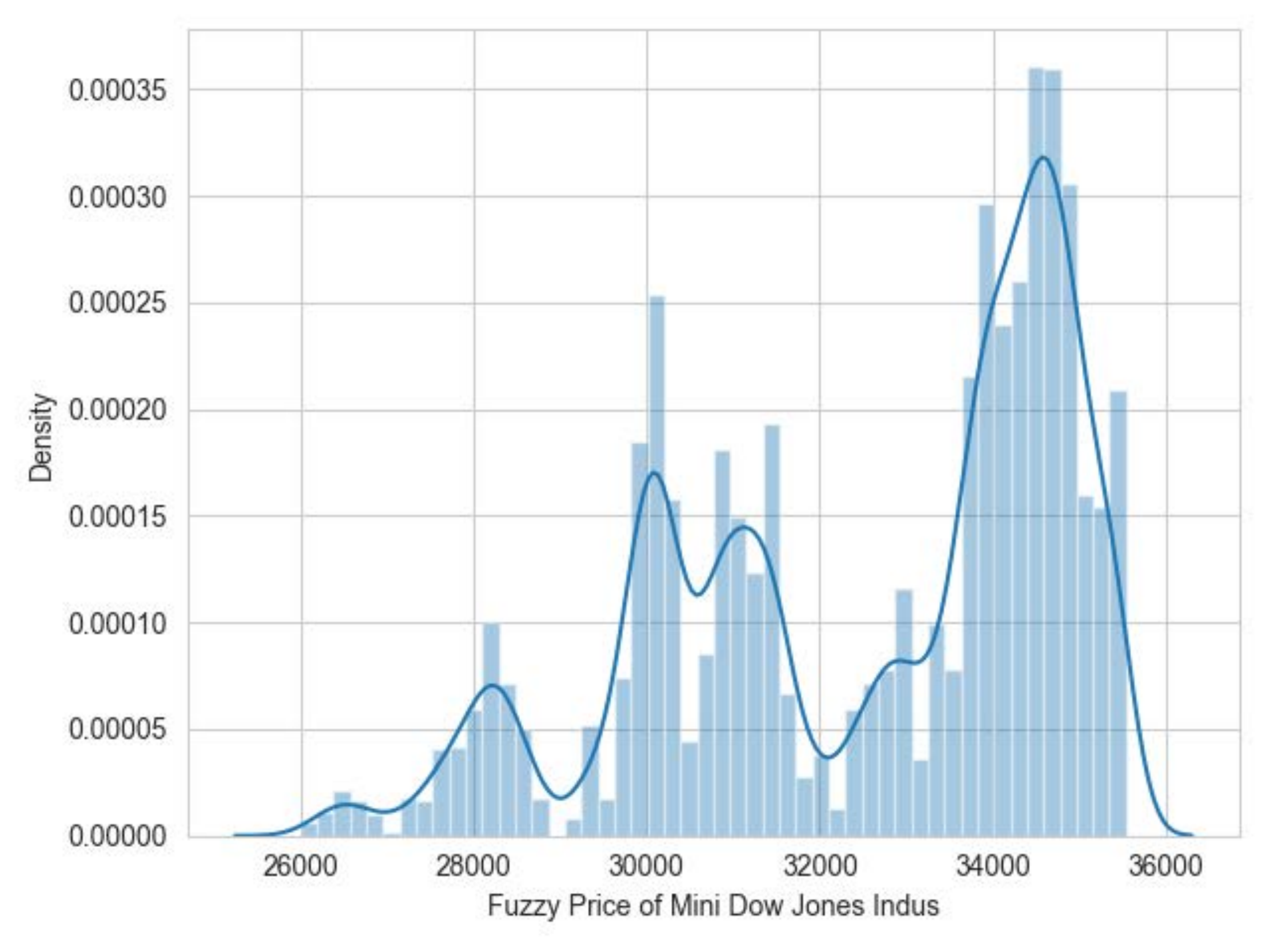}
\end{center}
\caption{Distribution plot for fuzzy price. \label{fig:second}}
\end{figure}

Figure 1 reports the distribution difference and change of fuzzy prices of mini-sized DJIA futures in each month of the year through the box chart. Obviously, the average price level in August 2021 is higher than that in other months. And the average level of the sample in October 2020 is the lowest. November 2020 has the largest price fluctuation. There were abnormal price changes in 7 months ( 10/2020, 12/2020, 1/2021, 2/2021, 5/2021, 6/2021, 7/2021 ), especially in September 2020 and October 2020.

As shown in Figure 2, the fuzzy price of mini-sized DJIA futures in one year is the most intensive in the range of $3400-3600$. Combined with figure 1, we could see that the prices after May $2021$ are mostly within this range, which shows that the overall fluctuation range of prices before May $2021$ is greater than that after May $2021$.

As mentioned in Subsection 3.2, the price change every 5 minutes is much smaller than the value of the price itself, so direct observation of the distribution of price change percentage would help us master a more precise information of sample dynamic change. Figure 3 is the histogram of fuzzy price change percentage of mini-sized DJIA futures. In the figure, the fuzzy price change statistics of mini-sized DJIA futures every 5 minutes can be positive or negative, mostly concentrated near $0$. The figure is not completely symmetrical and slightly skewed to the right, indicating that the number of prices experiencing rise and fall in the overall sample is similar, but the overall decline is less than the increase.

\begin{figure}[htp]
\begin{center}
\includegraphics[width=3in]{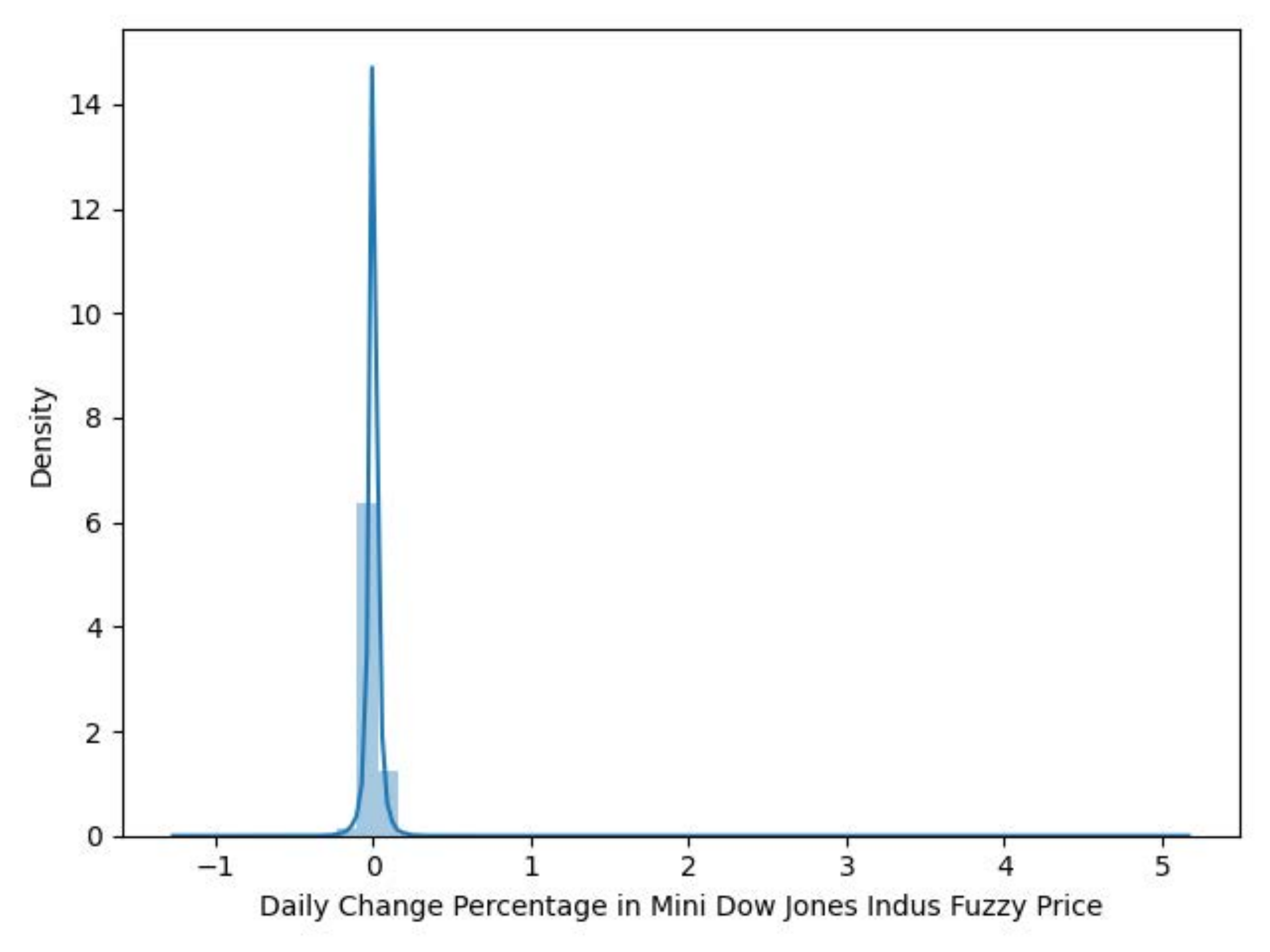}
\end{center}
\caption{Histogram for change percentage in fuzzy price. \label{fig:thrid}}
\end{figure}

\begin{figure}[htp]
\begin{center}
\includegraphics[width=3in]{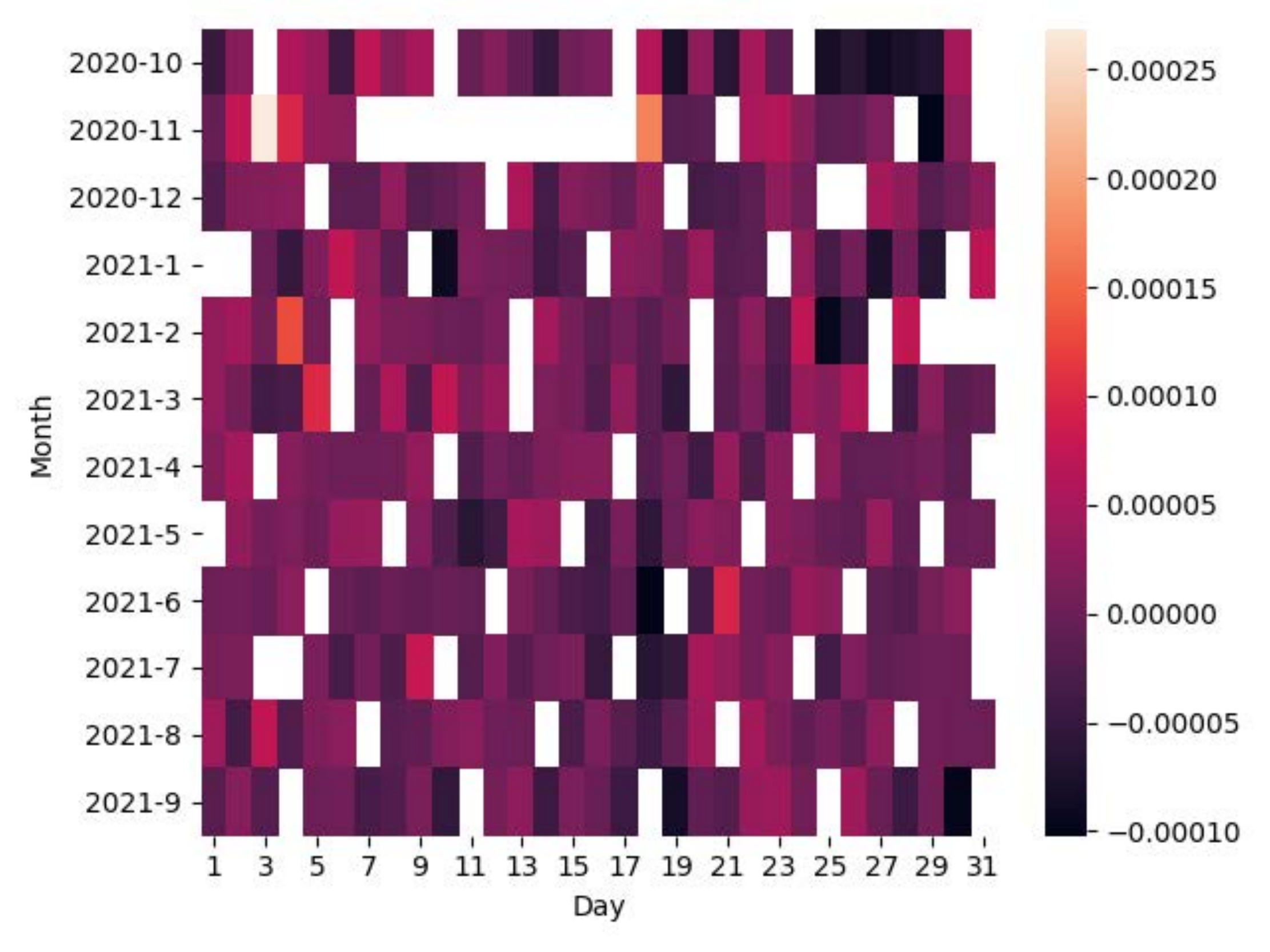}
\end{center}
\caption{Heatmap for the realized volatility. \label{fig:4th}}
\end{figure}

\begin{figure}[htp]
\begin{center}
\includegraphics[width=3in]{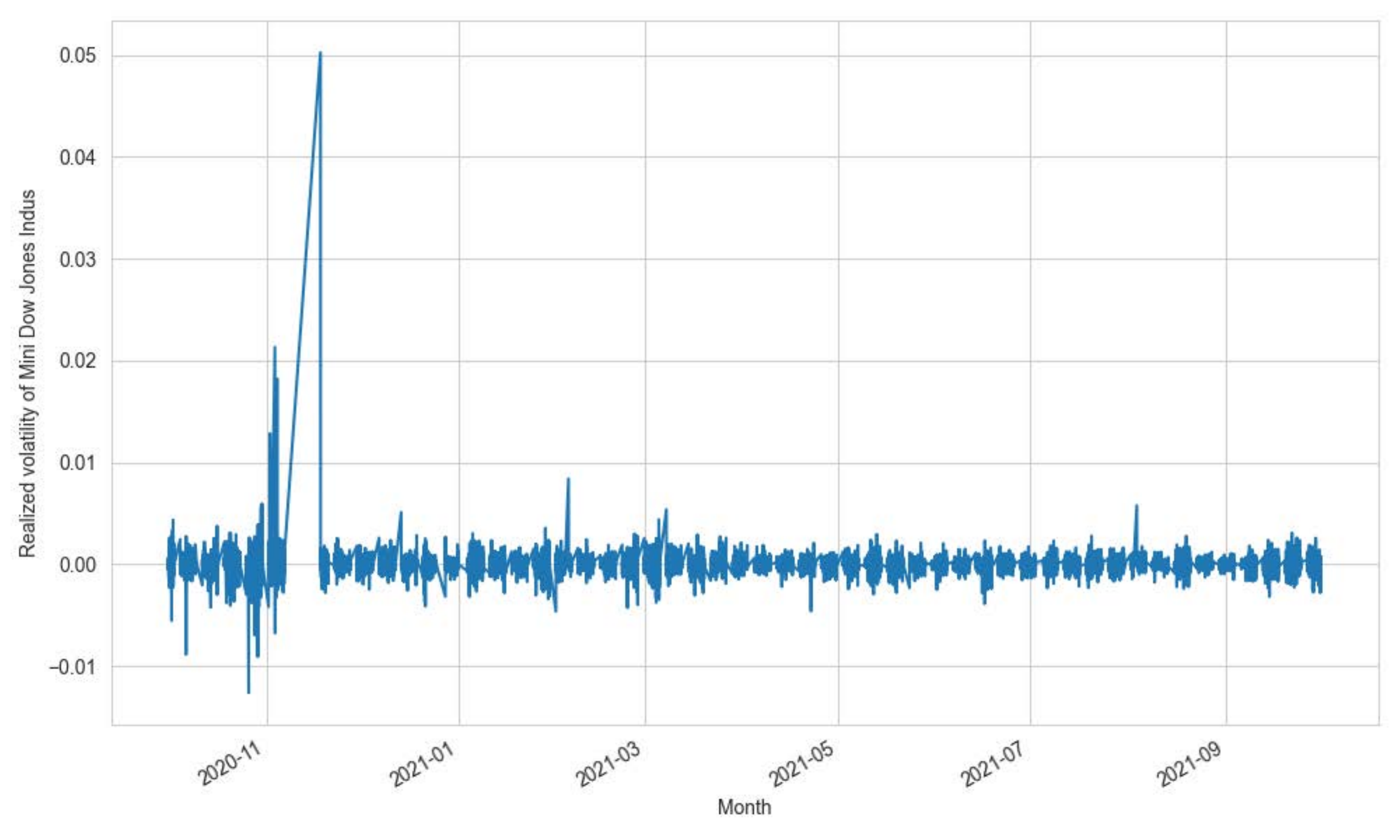}
\end{center}
\caption{Line Plot for the realized volatility. \label{fig:5th}}
\end{figure}

Now we observe the variation characteristics of realized volatility of mini-sized DJIA futures from the perspectives of numerical distribution and change trend. Figure 4 shows the realized volatility heatmap. By watching the black area and white area in the figure, the date with realized volatility change of more than $0.01\%$ could be identified. Obviously, November 2021 contains the largest number of volatile days. The realized volatility of each month has experienced frequent fluctuations. Figure 5 reports dynamic changes in realized volatility. The realized volatility of mini-sized DJIA futures has significant autocorrelation and volatility gathered. The sharp fluctuations are mainly concentrated near November 2021.

The analysis results in this section show that mini-sized DJIA futures always have volatility jumps of different amplitudes and frequencies in most periods of a year. For the sake of quantifying the jump characteristics in the fluctuation, we refer to the information of the data characteristics shown in the above chart and use the deep learning algorithms to estimate the parameters in the following section.

\section{Parameter Estimation}

\subsection{Classification Problem} Based on the data analysis results in Section 3, in this section, we estimate the value of $\theta$ in the new model in Section 2. We create a classification problem as follows.

1) Index the available fuzzy price of mini-sized DJIA futures in chronological order.

2) Group the data according to the price fluctuation characteristics analyzed in Section 3 and create a new data structure. We try to take the percentage change of fuzzy price for $50$ consecutive minutes as a row subset and stack it layer by layer to form a new fuzzy price matrix.

3) Consider the fluctuation range of fuzzy price change percentage of mini-sized DJIA futures, determine the threshold $K$, and search for the ``big jump". The time of each price data that is $K$ ``points" lower than the fuzzy price of the previous 5 minutes will be identified. ( if $K = 0.1\%$, record the time when the fuzzy price is $0.1\%$ lower than the previous one).

4) Create a target column $\theta$ for the new matrix. If there are at least two ``big jumps" in the next $50$ minutes, the value of the target column $\theta$ of the row is $1$. Otherwise, the value of the corresponding $\theta$ is $0$.

5) Neural network algorithms will be used for sample data set learning and parameter $\theta$ estimation.

After completing the steps above, assign $\theta$ the results of classification report. The value of $\theta$ could be directly used in formulas \eqref{(2.10)} and \eqref{(2.11)} as a deterministic component in the stochastic volatility model.

\subsection{Parameter Estimation Using Deep Learning}
In this subsection, we implement the steps in subsection 4.1 to calculate the results with reasonable accuracy $\theta$ to prove the operability of these steps. We run three common neural network algorithms ((a) standard neural network, (b) long short-term memory neural network and (c) LSTM network with batch normalizer (BN)) for classification calculation on python. The subset elements of each row in the new matrix are input parameters, and output parameter $\theta$ ($0$ or $1$) is the subset elements in the target column of the new matrix.

In a bid to report the applicability of the new model, the fluctuation rules of different time spans (intraday high-frequency data, daily data, weekly data and monthly data) should be taken into account. Referring to the sample characteristics in Section 3, We randomly select four classifiers from the preprocessed fuzzy price data for data training and data testing. The time periods of classifiers are

\noindent
$T1$: training time (Index): 09/26/2021\enspace 17:05\enspace (65039) to 09/27/2021 \enspace 12:05\enspace (65267); and testing time (Index): 09/27/2021\enspace 12:10 \enspace(65268) to 09/27/2021\enspace 14:55\enspace (65301);

\noindent
$T2$: training time (Index): 10/02/2020\enspace 0:00\enspace (357) to 10/05/2020 \enspace 23:55\enspace (902); and testing time (Index): 10/06/2020\enspace 0:00 \enspace(903) to 10/06/2020\enspace 23:55\enspace (1175);

\noindent
$T3$: training time (Index): 12/13/2020\enspace 17:05\enspace (10842) to 12/24/2020 \enspace 20:05\enspace (13258); and testing time (Index): 12/27/2020\enspace 17:05 \enspace(13259) to 12/31/2020\enspace 20:40\enspace (14352);

\noindent
$T4$: training time (Index): 05/02/2021\enspace 17:05\enspace (36922) to 07/30/2021 \enspace 16:00\enspace (54435); and testing time (Index): 08/01/2021\enspace 17:05 \enspace(54436) to 08/31/2021\enspace 23:55\enspace (60251).

Aiming at the different risk preference, threshold $K$ mentioned in the classification problem could be adjusted to satiate investors. The threshold $K$ of different sizes also helps market regulators identify and deal with the future volatility risk. The value of $K$, specifically, is the condition for us to retrieve the target fluctuation. The shorter the observation interval, the smaller the range of observed-value changing in a fixed time interval. Therefore, high sampling frequency is often best matched with small value of threshold $K$, which would identify more jumps. We use different $K$ sizes for the fuzzy price classifications of mini-sized DJIA futures. Table 2 reports the number of jumps that could be identified when different sizes of threshold $K$ are set in the classifications.

\begin{table}[htp]
\center

\vspace{-1.5em}
\caption{Time and index of the classifier \label{tab:tabtwo}}
\setlength\tabcolsep{1mm}{
\begin{tabular}{|c|c|c|c|c|c|}
\hline
 \small K size  & \small Jump number & \small \begin{tabular}[c]{@{}c@{}}Jump number \\ on T1\end{tabular} & \small \begin{tabular}[c]{@{}c@{}}Jump number \\ on T2\end{tabular} & \small \begin{tabular}[c]{@{}c@{}}Jump number \\ on T3\end{tabular} & \small \begin{tabular}[c]{@{}c@{}}Jump number \\ on T4\end{tabular} \\ \hline
KS=0.01 & 21566       & 87                                                           & 1026                                                         & 1106                                                         & 6871                                                         \\ \hline
KS=0.03 & 9626        & 30                                                           & 1084                                                         & 462                                                          & 2581                                                         \\ \hline
KS=0.05 & 4806        & 13                                                           & 112                                                          & 206                                                          & 1183                                                         \\ \hline
KS=0.1  & 1351        & 1                                                            & 49                                                           & 51                                                           & 259                                                          \\ \hline
KS=0.5  & 7           & 0                                                            & 3                                                            & 0                                                            & 0                                                            \\ \hline
KS=1    & 1           & 0                                                            & 0                                                            & 0                                                            & 0                                                            \\ \hline
\end{tabular}
}
\end{table}

Under different $K$ sizes, we use the three neural networks mentioned above to estimate the value of $\theta$. For the purpose of obtaining more accurate parameter estimation results, we set up four indicators to comprehensively evaluate the results of parameter prediction. ``support" reports the number of jump samples in the test set, ``precision" indicates the accuracy of the prediction results, ``recall" refers to prediction efficiency, and ``f1-score" shows the possibility of $\theta$ taking different values in the future. The classification reports are shown in Table 3-8.

\begin{table}[htp]
\center
\vspace{-1.5em}
\caption{Classification report in $T1$ $(KS=0.01)$ \label{tab:tabthree}}
\setlength\tabcolsep{1mm}{
\begin{tabular}{|c|c|c|c|c|c|c|c|c|}
\hline
KS=0.01 & \begin{tabular}[c]{@{}c@{}}precision \\ $\theta=0$\end{tabular} & \begin{tabular}[c]{@{}c@{}}recall \\ $\theta=0$\end{tabular} & \begin{tabular}[c]{@{}c@{}}f1-score \\ $\theta=0$\end{tabular} & \begin{tabular}[c]{@{}c@{}}support \\ $\theta=0$\end{tabular} & \begin{tabular}[c]{@{}c@{}}precision \\ $\theta=1$\end{tabular} & \begin{tabular}[c]{@{}c@{}}recall \\ $\theta=1$\end{tabular} & \begin{tabular}[c]{@{}c@{}}f1-score \\ $\theta=1$\end{tabular} & \begin{tabular}[c]{@{}c@{}}support \\ $\theta=1$\end{tabular} \\ \hline
(A)     & 0.00                                                            & 0.00                                                         & 0.00                                                           & \multirow{3}{*}{8}                                            & 0.76                                                            & 0.96                                                         & 0.85                                                           & \multirow{3}{*}{27}                                           \\ \cline{1-4} \cline{6-8}
(B)     & 0.00                                                            & 0.00                                                         & 0.00                                                           &                                                               & 0.75                                                            & 0.89                                                         & 0.81                                                           &                                                               \\ \cline{1-4} \cline{6-8}
(C)     & 0.00                                                            & 0.00                                                         & 0.00                                                           &                                                               & 0.76                                                            & 0.96                                                         & 0.85                                                           &                                                               \\ \hline
\end{tabular}
}
\end{table}
\begin{table}[htp]
\center
\vspace{-1.5em}
\caption{Classification report in $T1$ $(KS=0.05)$ \label{tab:tabfour}}
\setlength\tabcolsep{1mm}{
\begin{tabular}{|c|c|c|c|c|c|c|c|c|}
\hline
KS=0.05 & \begin{tabular}[c]{@{}c@{}}precision \\ $\theta=0$\end{tabular} & \begin{tabular}[c]{@{}c@{}}recall \\ $\theta=0$\end{tabular} & \begin{tabular}[c]{@{}c@{}}f1-score \\ $\theta=0$\end{tabular} & \begin{tabular}[c]{@{}c@{}}support \\ $\theta=0$\end{tabular} & \begin{tabular}[c]{@{}c@{}}precision \\ $\theta=1$\end{tabular} & \begin{tabular}[c]{@{}c@{}}recall \\ $\theta=1$\end{tabular} & \begin{tabular}[c]{@{}c@{}}f1-score \\ $\theta=1$\end{tabular} & \begin{tabular}[c]{@{}c@{}}support \\ $\theta=1$\end{tabular} \\ \hline
(A)     & 0.54                                                            & 1.00                                                         & 0.70                                                           & \multirow{3}{*}{19}                                           & 0.00                                                            & 0.00                                                         & 0.00                                                           & \multirow{3}{*}{16}                                           \\ \cline{1-4} \cline{6-8}
(B)     & 0.54                                                            & 1.00                                                         & 0.70                                                           &                                                               & 0.00                                                            & 0.00                                                         & 0.00                                                           &                                                               \\ \cline{1-4} \cline{6-8}
(C)     & 0.53                                                            & 0.95                                                         & 0.68                                                           &                                                               & 0.00                                                            & 0.00                                                         & 0.00                                                           &                                                               \\ \hline
\end{tabular}
}
\end{table}
\begin{table}[htp]
\center
\vspace{-1.5em}
\caption{Classification report in $T2$ $(KS=0.01)$\label{tab:tabfive}}
\setlength\tabcolsep{1mm}{
\begin{tabular}{|c|c|c|c|c|c|c|c|c|}
\hline
KS=0.01 & \begin{tabular}[c]{@{}c@{}}precision \\ $\theta=0$\end{tabular} & \begin{tabular}[c]{@{}c@{}}recall \\ $\theta=0$\end{tabular} & \begin{tabular}[c]{@{}c@{}}f1-score \\ $\theta=0$\end{tabular} & \begin{tabular}[c]{@{}c@{}}support \\ $\theta=0$\end{tabular} & \begin{tabular}[c]{@{}c@{}}precision \\ $\theta=1$\end{tabular} & \begin{tabular}[c]{@{}c@{}}recall \\ $\theta=1$\end{tabular} & \begin{tabular}[c]{@{}c@{}}f1-score \\ $\theta=1$\end{tabular} & \begin{tabular}[c]{@{}c@{}}support \\ $\theta=1$\end{tabular} \\ \hline
(A)     & 0.25                                                            & 0.03                                                         & 0.06                                                           & \multirow{3}{*}{29}                                           & 0.90                                                            & 0.99                                                         & 0.94                                                           & \multirow{3}{*}{245}                                          \\ \cline{1-4} \cline{6-8}
(B)     & 0.00                                                            & 0.00                                                         & 0.00                                                           &                                                               & 0.89                                                            & 0.99                                                         & 0.94                                                           &                                                               \\ \cline{1-4} \cline{6-8}
(C)     & 0.00                                                            & 0.00                                                         & 0.00                                                           &                                                               & 0.89                                                            & 0.96                                                         & 0.93                                                           &                                                               \\ \hline
\end{tabular}
}
\end{table}
\begin{table}[htp]
\center
\vspace{-1.5em}
\caption{Classification report in $T2$ $(KS=0.03)$\label{tab:tabsix}}
\setlength\tabcolsep{1mm}{
\begin{tabular}{|c|c|c|c|c|c|c|c|c|}
\hline
KS=0.03 & \begin{tabular}[c]{@{}c@{}}precision \\ $\theta=0$\end{tabular} & \begin{tabular}[c]{@{}c@{}}recall \\ $\theta=0$\end{tabular} & \begin{tabular}[c]{@{}c@{}}f1-score \\ $\theta=0$\end{tabular} & \begin{tabular}[c]{@{}c@{}}support \\ $\theta=0$\end{tabular} & \begin{tabular}[c]{@{}c@{}}precision \\ $\theta=1$\end{tabular} & \begin{tabular}[c]{@{}c@{}}recall \\ $\theta=1$\end{tabular} & \begin{tabular}[c]{@{}c@{}}f1-score \\ $\theta=1$\end{tabular} & \begin{tabular}[c]{@{}c@{}}support \\ $\theta=1$\end{tabular} \\ \hline
(A)     & 0.30                                                            & 0.26                                                         & 0.28                                                           & \multirow{3}{*}{84}                                           & 0.69                                                            & 0.73                                                         & 0.71                                                           & \multirow{3}{*}{190}                                          \\ \cline{1-4} \cline{6-8}
(B)     & 0.40                                                            & 0.36                                                         & 0.38                                                           &                                                               & 0.73                                                            & 0.76                                                         & 0.75                                                        &                                                               \\ \cline{1-4} \cline{6-8}
(C)     & 0.33                                                            & 0.15                                                         & 0.21                                                           &                                                               & 0.70                                                            & 0.86                                                         & 0.77                                                          &                                                               \\ \hline
\end{tabular}
}
\end{table}

The recognition results of the jumps number in historical price by use the three neural networks are always consistent. It shows that the three selected algorithms are all effective and accurate at the jump recognition stage.

Comparing table 3 and 4, we found that with the increasing of $K$, the numbers of response in $T1$ to $\theta=0$ increases significantly and that to $\theta=1$ gradually decreases. This result is in line with our expectation. In the same time interval, the more jumps could be captured with the lower threshold of jump recognition, and it is greater possibility of $\theta=1$, which means the stronger fuzzy random process $\tilde{Z}_{\lambda t}^{(b)}$ will be used to characterize the realized volatility dynamics.
\begin{table}[htp]
\center
\vspace{-1.5em}
\caption{Classification report in $T3$ $(KS=0.05)$\label{tab:tabseveen}}
\setlength\tabcolsep{1mm}{
\begin{tabular}{|c|c|c|c|c|c|c|c|c|}
\hline
KS=0.05 & \begin{tabular}[c]{@{}c@{}}precision \\ $\theta=0$\end{tabular} & \begin{tabular}[c]{@{}c@{}}recall \\ $\theta=0$\end{tabular} & \begin{tabular}[c]{@{}c@{}}f1-score \\ $\theta=0$\end{tabular} & \begin{tabular}[c]{@{}c@{}}support \\ $\theta=0$\end{tabular} & \begin{tabular}[c]{@{}c@{}}precision \\ $\theta=1$\end{tabular} & \begin{tabular}[c]{@{}c@{}}recall \\ $\theta=1$\end{tabular} & \begin{tabular}[c]{@{}c@{}}f1-score \\ $\theta=1$\end{tabular} & \begin{tabular}[c]{@{}c@{}}support \\ $\theta=1$\end{tabular} \\ \hline
(A)     & 0.96                                                            & 094                                                          & 0.95                                                           & \multirow{3}{*}{1049}                                         & 0.11                                                            & 0.15                                                         & 0.12                                                           & \multirow{3}{*}{46}                                           \\ \cline{1-4} \cline{6-8}
(B)     & 0.96                                                            & 0.92                                                         & 0.94                                                           &                                                               & 0.11                                                            & 0.22                                                         & 0.15                                                         &                                                               \\ \cline{1-4} \cline{6-8}
(C)     & 0.96                                                            & 0.95                                                         & 0.95                                                           &                                                               & 0.10                                                            & 0.13                                                         & 0.11                                                        &                                                               \\ \hline
\end{tabular}
}
\end{table}

\begin{table}[htp]
\center
\vspace{-1.5em}
\caption{Classification report in $T4$ $(KS=0.1)$\label{tab:tabeight}}
\setlength\tabcolsep{1mm}{
\begin{tabular}{|c|c|c|c|c|c|c|c|c|}
\hline
KS=0.05 & \begin{tabular}[c]{@{}c@{}}precision \\ $\theta=0$\end{tabular} & \begin{tabular}[c]{@{}c@{}}recall \\ $\theta=0$\end{tabular} & \begin{tabular}[c]{@{}c@{}}f1-score \\ $\theta=0$\end{tabular} & \begin{tabular}[c]{@{}c@{}}support \\ $\theta=0$\end{tabular} & \begin{tabular}[c]{@{}c@{}}precision \\ $\theta=1$\end{tabular} & \begin{tabular}[c]{@{}c@{}}recall \\ $\theta=1$\end{tabular} & \begin{tabular}[c]{@{}c@{}}f1-score \\ $\theta=1$\end{tabular} & \begin{tabular}[c]{@{}c@{}}support \\ $\theta=1$\end{tabular} \\ \hline
(A)    & 0.99                                                            & 1.00                                                         & 0.99                                                           & \multirow{3}{*}{5747}                                         & 0.00                                                            & 0.00                                                         & 0.00                                                           & \multirow{3}{*}{70}                                           \\ \cline{1-4} \cline{6-8}
(B)    & 0.99                                                            & 0.99                                                         & 0.99                                                           &                                                               & 0.07                                                            & 0.06                                                         & 0.06                                                          &                                                               \\ \cline{1-4} \cline{6-8}
(C)    & 0.99                                                            & 0.99                                                         & 0.99                                                           &                                                               & 0.04                                                            & 0.03                                                         & 0.03                                                           &                                                               \\ \hline
\end{tabular}
}
\end{table}

The classification results in $T1$ show that it is more likely that the value of $\theta$ in the test set is $1$ when $KS = 0.01$. And that's the fact. When $KS = 0.05$, all three algorithms suggest that $\theta=0$. These results show that the value of $\theta$ is affected by the size of threshold $K$. In other words, our new model could track and estimate jumps with different jump size and the jump intensity. It is flexible and suitable for investors and regulators with different purposes to grasp the change law of the market changes.

A similar conclusion is drawn in the comparison between table 5 and 6. For $KS = 0.01$, the numbers of response to $\theta$ =0 is fewer than that to $KS = 0.03$. The numbers of response to $\theta$ =1 is negatively correlated with the value of K. The results in tables 5 and 6 show that 1 is considered to be an appropriate value of $\theta$ for $KS \leq 0.03$ in $T2$. And the same results are reported in table 7 and 8. Where $\theta$ is suggested to be $0$ for $KS = 0.05$ in T3 and $KS = 0.1$ in $T4$. This means that, it is suitable for a weaker fuzzy stochastic process $\tilde{Z}_{\lambda t}$ used to characterize realized volatility dynamics.

Comparing the ``support" in table 3 and 5, we find that the more jumps could be recognized with the increase of time span $T$. This rule would also be found by comparing the results in table 4 and 7.

In conclusion, both $K$ and $T$ can directly affect the number of recognized jumps, and then affect the value of $\theta$. The effect of value of $K$ on number of identifiable jumps was positive and significant, and the effect of $T$ length on number of identifiable jumps was negative and significant.

\section{Conclusion}
\label{sec:con}
The research on volatility estimation plays a vital role in asset pricing and market risk prediction. How to improve existing volatility prediction model to make it more suitable for the complex financial market environment has been paid attention by the academic and industry. In this paper, we contributed to the existing research on stochastic volatility estimation in two ways. Firstly, we considered the fuzziness and randomness of the price process and used fuzzy random variables to replace the random variables in the classical BN-S model, which avoids the influence of microstructure noise in high-frequency data to a certain extent and improves the description accuracy of historical data. Secondly, referring to the research results on OU process superposition, we discussed the advantages of superposition fuzzy random process, and proposed new parameter to overcome the lack of long-term dependence of BN-S model.

We selected the trading price data of mini-sized DJIA futures within one year as the sample, analyzed the existence and distribution characteristics of jumps in the price process under the sampling frequency of 5 minutes, and realized the dynamic price volatility estimation under uncertain environment with fuzziness and randomness. Using the deep learning algorithms, we recognized the jumps of different intensities and their distribution in the historical data, and capture the deterministic components, which would be used for volatility prediction. Through the statistical analysis of sample data, the operability and effectiveness of the proposed generalized new BN-S model were illustrated. As we know, the new model described in this paper is also applicable to forecast volatility of many other financial assets. Furthermore, the new model offers promising avenues for pricing, regulation and risk control.

There is room for further improvement in this paper. Compared with the deep learning algorithms used in this paper, the new big data algorithm may improve the parameter estimation results in the classification problem. The optimal sampling frequency of different financial assets in the new model can’t be determined. These limitations deserve to be considered and improved in future research.

\section*{Acknowledgments}
This work is jointly supported
by the National Natural Science Foundation of China under Nos. 11662001 and 11771105, the Science Foundation of Guangxi Province under
Nos. 2017GXNSFFA198012 and 2018GXNSFAA138177.

\bibliography{sn-bibliography}


\end{document}